\documentclass[conference]{IEEEtran}
\IEEEoverridecommandlockouts
\usepackage{csquotes}
\usepackage{cite}
\usepackage{amssymb}
\usepackage{graphicx}
\usepackage{epstopdf}
\usepackage{amsmath}
\usepackage{amsthm}
\usepackage[usenames, dvipsnames]{color}
\usepackage{algorithm}
\usepackage{algpseudocode}
\usepackage{multicol}
\usepackage{lipsum}
\usepackage{mwe}
\usepackage{hyperref}

{}
{}
{}
\begin{document}

\title{\LARGE \bf Adaptive Intelligent Secondary Control of Microgrids Using a Biologically-Inspired Reinforcement Learning}

\author{Mohammad Jafari, Vahid Sarfi, Amir Ghasemkhani, Hanif Livani, Lei Yang, and Hao Xu
\thanks{*This work is partially supported by NSF award, \# 1723814.}
\thanks{M.~Jafari is with the Department of Applied Mathematics, Jack Baskin School of Engineering, University of California, Santa Cruz, CA 95064.
        {\tt\small mjafari1@ucsc.edu}}
\thanks{V.~Sarfi, H.~Livani and H.~Xu are with the Department of Electrical and Biomedical Engineering, University of Nevada, Reno, NV 89557-0260.
        {\tt\small vsarfi@nevada.unr.edu,[hlivani, haoxu]@unr.edu}}%
\thanks{A.~Ghasemkhani and L.~Yang are with the Department of Computer Science and Engineering, University of Nevada, Reno, NV 89557-0260.
        {\tt\small aghasemkhani@nevada.unr.edu, leiy@unr.edu}}
}

\maketitle

\begin{abstract}

In this paper, a biologically-inspired adaptive intelligent secondary controller is developed for microgrids to tackle system dynamics uncertainties, faults, and/or disturbances. The developed adaptive biologically-inspired controller adopts a novel computational model of emotional learning in mammalian limbic system. The learning capability of the proposed biologically-inspired intelligent controller makes it a promising approach to deal with the power system non-linear and volatile dynamics without increasing the controller complexity, and maintain the voltage and frequency stabilities by using an efficient reference tracking mechanism. The performance of the proposed intelligent secondary controller is validated in terms of the voltage and frequency absolute errors in the simulated microgrid. Simulation results highlight the efficiency and robustness of the proposed intelligent controller under the fault conditions and different system uncertainties compared to other benchmark controllers.
\end{abstract}

\begin{IEEEkeywords}
Intelligent Secondary Controller, Microgrids, BELBIC, Emotional Learning
\end{IEEEkeywords}

\section{Introduction}

\subsection{Motivation}

Microgrids (MGs) with their corresponding control systems are independent distribution power systems which provide guaranteed power quality for various loads~\cite{bidram2014distributed}. MGs are operational in both islanded and grid connected modes. This operational flexibility provides an opportunity for the scalable integration of local generators including distributed energy resources (DERs) into power system. However, integrating DERs puts forth stability and operational issues for the MGs' operators. To this end, MG energy management system (EMS) needs to incorporate controlling schemes for MGs in order to maintain the stability of the system and address the operational issues in both steady and faulty states. Additionally, these controlling schemes should be efficient and robust enough to account for various system uncertainties in different operational modes and states~\cite{bevrani2012intelligent,lasseter2011certs}.

A hierarchical control scheme is applied to address the challenges of MGs’ operation in both islanded and grid-connected modes~\cite{bidram2014distributed}. A hierarchical control scheme comprises three levels: primary, secondary and tertiary control levels. Primary control level is responsible for generating fast control responses including inner voltage/current and power sharing control signals to maintain the voltage and frequency stability. Secondary control acts over primary control by sending compensation references to restore the voltage and frequency deviations to the nominal values. The highest level of control, the so-called tertiary level is required to specify the optimal set-points for operation of the generation resources by considering the power system requirements~\cite{mohamed2017hierarchical}. Reviewing the related works in MGs control schemes shows that most of the state-of-art methods require detailed information about the system dynamics. In this sense, developing a model-free adaptive controller becomes of practical importance due to nonlinear and complicated nature of the DERs’ dynamics in MGs.

\subsection{Related works} 

Previous works have addressed the challenges in MG operation by proposing several control methods. The autonomous operation of MGs during the transition to the islanded mode have been proposed in~\cite{katiraei2005micro}. Nonlinear heterogeneous dynamics of
DERs has been transformed into linear dynamics using an input-output feedback linearizion approach in~\cite{bidram2014distributed} to design a secondary controller for MGs.
Recently, authors in~\cite{bevrani2012intelligent} have proposed an adaptive PI based frequency controller for MGs by leveraging a combination of the fuzzy logic and the particle swarm optimization techniques to improve the conventional PI controller performance against dynamical changes. An intelligent pinning based cooperative secondary control of DERs for an islanded MG has been developed in~\cite{manaffam2017intelligent}. Moreover, decentralized and distributed secondary controllers have been investigated for MGs in islanded mode in~\cite{lou2017decentralised} and \cite{li2018distributed}, respectively. All of these controllers propose an efficient performance with respect to the changes in the system operating conditions. However, they require the detailed information about the system dynamics to update the MG control parameters in real time. To tackle this problem, a model-free secondary controller
has been studied in~\cite{bevrani2013intelligent}. In this sense, an adaptive neuro-fuzzy inference system (ANFIS) method has been proposed for simultaneous
voltage and frequency control in an islanded MG. Although the suggested controller performs well with respect to tracking the changes in the normal operating conditions,
the proposed control scheme is trained by a desired I/O data set offline
which is not appropriate since it needs to be applied to an actual
system. Hence, there is a pressing need to develop control strategies with less
dependency on the full knowledge of the system dynamics which
can adjust MG operating parameters online.

In the past few years, diverse complex problems have been successfully solved by extensively employing the intelligent techniques~\cite{jafari2013attitude,klecker2017robust, yin2017artificial, jafari2018power}. Brain Emotional Learning Based Intelligent Controller (BELBIC)~\cite{lucas2004introducing}, is a biologically-inspired intelligent model-free controller which can be successfully implemented into complex problems by assigning appropriate functions to the Sensory Inputs (SI) and Emotional Signal (ES) which are the two main inputs of the BELBIC model~\cite{jafari2017mas,kim2017brain}. Finally, BELBIC has shown promising performance even when the model dynamics are fully or partially unknown and/or there exist noise and system uncertainty~\cite{lucas2004introducing}.

\subsection{Main Contributions}
In this paper, we focus on two issues of the MGs secondary controller design, i.e., the effect of the system dynamics in the design process while they are fully or partially unknown, and the robustness of the controller with respect to the system uncertainties in different operational conditions. Our main contributions are summarized as follows:
\begin{itemize}
\item A model-free adaptive intelligent secondary controller is developed for voltage and frequency stabilization in MGs with system uncertainties and disturbances. This controller not only is able to track the changes in different operating conditions, but also updates the required controlling commands in real time.
\item The intelligent secondary controller is designed by leveraging the concept of brain emotional learning mechanism. BELBIC is a model-free controller which performs well in presence of system noise an uncertainties. Moreover, it has a low computational complexity which makes it a promising method in a real-time application. The proposed controller not only reduces the system complexity, but also provides a controller with multi-objective properties (i.e., control effort optimization, uncertainty handling, and noise/disturbance rejection).
\item Lyapunov analysis is provided to show the convergence of the proposed intelligent controller. The learning capability of the proposed approach is validated for stabilizing the voltage and frequency of the MGs. In order to demonstrate the effectiveness of the proposed approach, a comparison between the proposed approach and both conventional PID controller and an intelligent controller based on neural network is provided. The results show superior performance of the BELBIC in terms of robustness and adaptivity.
\item Additionally, since in the proposed control framework, the control inputs related to the individual agent is computed without the knowledge of the states of the other agents, therefore, it can be utilized for decentralized control of the multiple agents. However, for brevity, we applied the proposed secondary controller to an MG consisting of two synchronous generators, while the secondary controller is only providing the control inputs for the first generator without the knowledge of the states of the second generator. This is a good example for showing the decentralized capability of the proposed controller. The more rigorous study with focusing on the decentralized control of multiple agents will be provided in our future works.  
\end{itemize}

\section{Brain Emotional Learning-Based Intelligent Controller}
\label{BELBIC}

\begin{figure}[t!]
\vspace{-10pt}
   \centering
   \includegraphics[width=0.55\columnwidth]{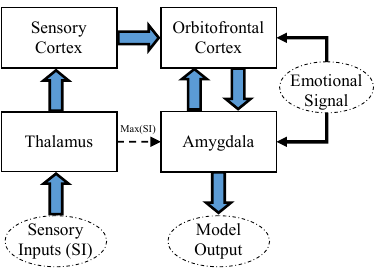}
   \vspace{-10pt}
   \caption{Computational model of emotional learning.}
   \label{mor}
   \vspace{0pt}
\end{figure}

BELBIC is a neurobiologically-inspired intelligent model-free controller which takes advantage of the mathematical model of the emotional learning in the mammalian limbic system introduced in~\cite{moren2000computational}. This model (depicted in Fig.~\ref{mor}), has \textit{Amygdala}, and \textit{Orbitofrontal Cortex} as its primary components. According to the mammalian limbic system, the role of Amygdala is immediate learning, while Orbitofrontal Cortex plays an inhibitory role to avoid any inappropriate learning happening in the Amygdala. Furthermore, BELBIC model has \textit{Sensory Inputs} ($SI$), and \textit{Emotional Signal} ($ES$) as its two external inputs. 

Amygdala outputs are calculated by the summation of all its corresponding nodes, where the output of each node is described as equation~\eqref{amyg} and the equation~\eqref{amyg_l} is employed for updating its weights (i.e., $V_i$).
\begin{align}
\label{amyg}
A_i = V_i\times SI_i
\end{align}
\begin{align}
\label{amyg_l}
\Delta V_i = K_v \times SI_i \times \max\left( 0, ES - \sum_{i} A_i\right)
\end{align}
where, $K_v$ is the learning rate.

The maximum of all $SIs$ is another input considered in the model. This signal (i.e., $A_{th}$), which is directly sent from the Thalamus to the Amygdala, is defined as:
\begin{align}
\label{maxamyg}
A_{th} = V_{th} \times \max\left(SI_i\right)
\end{align}
where $V_{th}$ is the weight and the corresponding update law is the same as Equation~(\ref{amyg_l}).

To calculate the Orbitofrontal Cortex outputs, all its corresponding nodes are added together, where the equation~\eqref{OFC} is describing the output of each node and for updating its weights (i.e., $W_i$) the equation~\eqref{OFC_l} is employed.
\begin{align}
\label{OFC}
OC_i = W_i \times SI_i
\end{align}
\begin{align}
\label{OFC_l}
\Delta W_i = K_w \times SI_i \times \left( MO - ES \right)
\end{align}
where, $K_w$ is the learning rate. The output of the BELBIC model ($MO$) is calculated by the difference between Amygdala outputs ($A_i$) and Orbitofrontal Cortex outputs ($OC_i$) and can be obtained as follows:
\begin{align}
\label{bel_out}
MO = \sum_{i} \left[ A_i - OC_i \right],
\end{align}
where in all the equations, $i$ is the number of sensory inputs.

In order to tune the parameters of BELBIC model, several techniques have been adopted. 
In this paper a heuristic approach is utilized for tuning the BELBIC parameters, in order to significantly reduce the computational complexity of the overall system. In other words, this algorithm provides boundaries for these parameters, such that the MGs estimated weights of Amygdala and Orbitofrontal Cortex converge to desired targets asymptotically (see the results obtained in Theorem 1 in \cite{JafariBELBIC2018}). 
\subsection{Objectives}
\label{obj}
The main objective of this paper is to design two control signals $u_1$ and $u_2$ (i.e., $U_{sec}$ and $T_{sec}$), for simultaneously stabilizing the frequency and the voltage of the MGs by proposing a biologically-inspired adaptive intelligent secondary controller. The proposed  secondary level control is designed to stabilize both frequency and voltage signals in the events of disturbances by leveraging the computational model of emotional learning in mammalian limbic system (i.e., BELBIC) introduced in Section~\ref{BELBIC}, and based on the hierarchical control structure of MGs.

\section{BELBIC-based Intelligent Secondary Control for Microgrids}  \label{section:BELBIC_MG}

\subsection{System Design}
\label{subsection:Bel_design}

\begin{figure}[t!]
\vspace{-10pt}
    \centering
  \includegraphics[width=0.8\columnwidth]{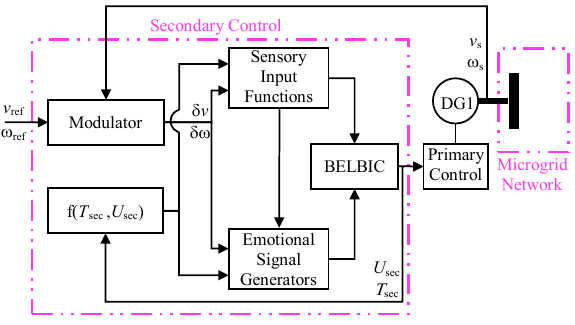}
  \vspace{-12pt}
    \caption{Proposed control architecture.}
    \label{bell}
    \vspace{0pt}
\end{figure}

The BELBIC architecture implemented in this work is shown in Fig.~\ref{bell} where it demonstrates a closed loop configuration. This architecture implicitly demonstrates the overall emotional learning based control concept, which consists of the action selection mechanism, the critic, and the learning algorithm~\cite{lucas2004introducing}. The proposed architecture consists of the following blocks: (i)~BELBIC, (ii)~Sensory input functions, (iii)~Emotional signal generators, (iv)~Modulator, (v)~Function of control efforts (i.e., $f(T_{sec},U_{sec})$), (vi)~Primary control, and finally (vii)~a block for the $DG_1$. 

\subsection{Emotional Signal and Sensory Input Development}
\label{subsection:ES_dev}
Generally, from the artificial intelligence (AI) point of view, BELBIC is considered as an action selection technique. In this technique, the sensory input ($SI$) and the emotional signal ($ES$) are responsible for producing the action. The following functions can represent $SI$ and $ES$ in their general forms.
\begin{align}
\label{SI-m}
SI = \mathcal{F}\left( r, e, u, y \right)
\end{align}
\begin{align}
\label{ES-m}
ES = \mathcal{G}\left(r, e, u, y \right)
\end{align}
where in our case,  $r=\{\nu_{ref}, \omega_{ref}\}$ is the set of system inputs which corresponds to the voltage and frequency references, $e=\{\delta \nu, \delta\omega\}$ is the set of system frequency and voltage errors for $SM_1$, $u=\{U_{sec}, T_{sec}\}$ is the set of the control inputs which contains the control inputs for $SM_1$, and $y=\{\omega_s, \nu_s\}$ is the set of system outputs consisting of the measured system frequency and voltage for $SM_1$. 

By choosing the appropriate $ES$ for the MG systems, different control objectives, such as simultaneously stabilizing the frequency and voltage are achieved. In this sense, frequency and voltage references are dynamically generated for the primary control to obtain a better tracking performance of the MGs. Furthermore, it is possible to incorporate model uncertainties, such as network configuration and parameters, and other operational issues such as fault and/or disturbance events, in the proposed secondary control scheme.

In this paper, the proposed biologically-inspired technique will focus on improving: (i)~reference tracking performance, (ii)~model uncertainty handling, and (iii)~disturbance rejection. Therefore, to accomplish these objectives, a model-free adaptive intelligent secondary control for voltage and frequency stabilization in MGs with following characteristics is proposed. Here, the corresponding $SI_l$ and $ES_l$ for each of the control inputs (i.e., \{$u_1$,$u_2$\}), are designed as:
\begin{align}
\label{SI}
SI_l = K_{l,1} e_l + K_{l,2} \int e_l.dt + K_{l,3} \frac{de_l}{dt}
\end{align}
\begin{align}
\label{ES}
ES_l = K_{l,4} |u_l| + K_{l,5} e_l + K_{l,6} \int e_l.dt + K_{l,7} \frac{de_l}{dt}
\end{align}
where $l=\{1,2\}$ is an index representing the control inputs, $K_{l,1}$, $K_{l,2}$, $K_{l,3}$, $K_{l,4}$, $K_{l,5}$, $K_{l,6}$, and $K_{l,7}$ are positive gains. In general, assigning different values to these positive gains will change the $ES$ impacts on the system behavior. These gains are assigned differently for each one of the control inputs (i.e., $u_l$, $l = 1,2$) of the system.

\subsection{Learning-based Secondary Control}
\label{subsection:Learning_track}

Multiple performance considerations have to be taken into account simultaneously in designing secondary controllers for MGs. Therefore, utilization of methodologies with multi-objective capability, such as biologically-inspired learning-based techniques, is of paramount importance. To this end, motivated by the computational model of emotional learning in the mammals' limbic system, i.e., BELBIC, a model-free adaptive intelligent secondary control for voltage and frequency stabilization of MGs is proposed. This secondary controller is designed in such a way that not only it takes into account the system uncertainties and disturbances, and also it is a suitable controller for real-time implementation.

Considering the emotional learning model formulations in~\ref {BELBIC}, and equations~\eqref{SI}-\eqref{ES}, the BELBIC-inspired secondary control strategy for voltage and frequency stabilization in MGs is defined as
\begin{align}
\label{u_gamman}
\nonumber u_l = &V_l\times SI_l - W_l\times SI_l \\
\nonumber = & V_l\times \left(K_{l,1} e_l + K_{l,2} \int e_l.dt + K_{l,3} \frac{de_l}{dt} \right) \\
   &- W_l\times \left(K_{l,1} e_l + K_{l,2} \int e_l.dt + K_{l,3} \frac{de_l}{dt} \right)
\end{align}
here, $l=1,2$ makes reference to each control input. 

A detailed stability and convergence analyses of the weights of the Amygdala ($V_l$) and the Orbitofrontal Cortex are relegated to \cite{JafariBELBIC2018} due to page limitation.

\section{Simulation Results and Discussion} 
\label{section:simulation_results}
\begin{figure*}[htb!]
\begin{center}$
\begin{array}{cccc}
\includegraphics[width=0.5\columnwidth]{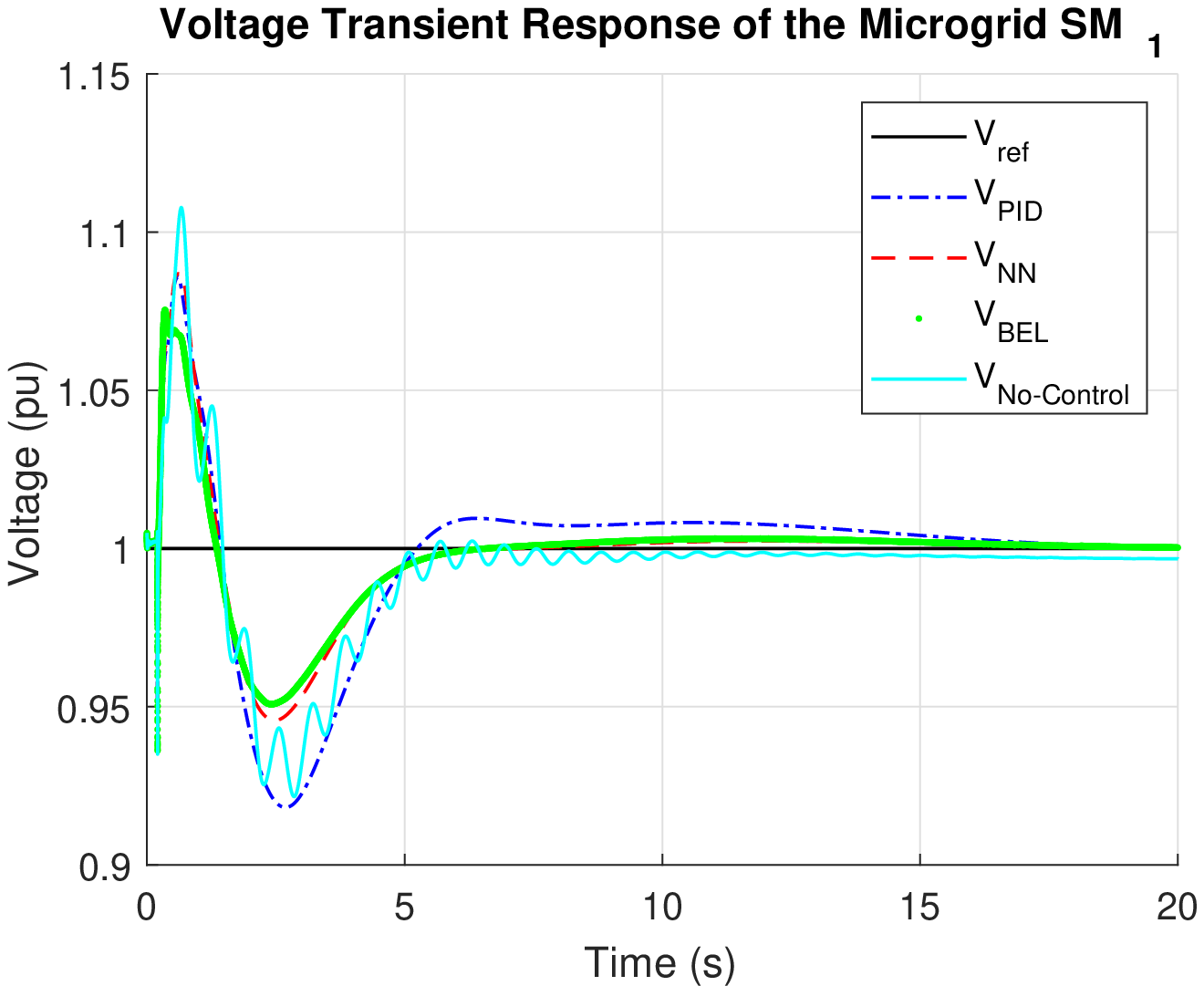} 
\includegraphics[width=0.5\columnwidth]{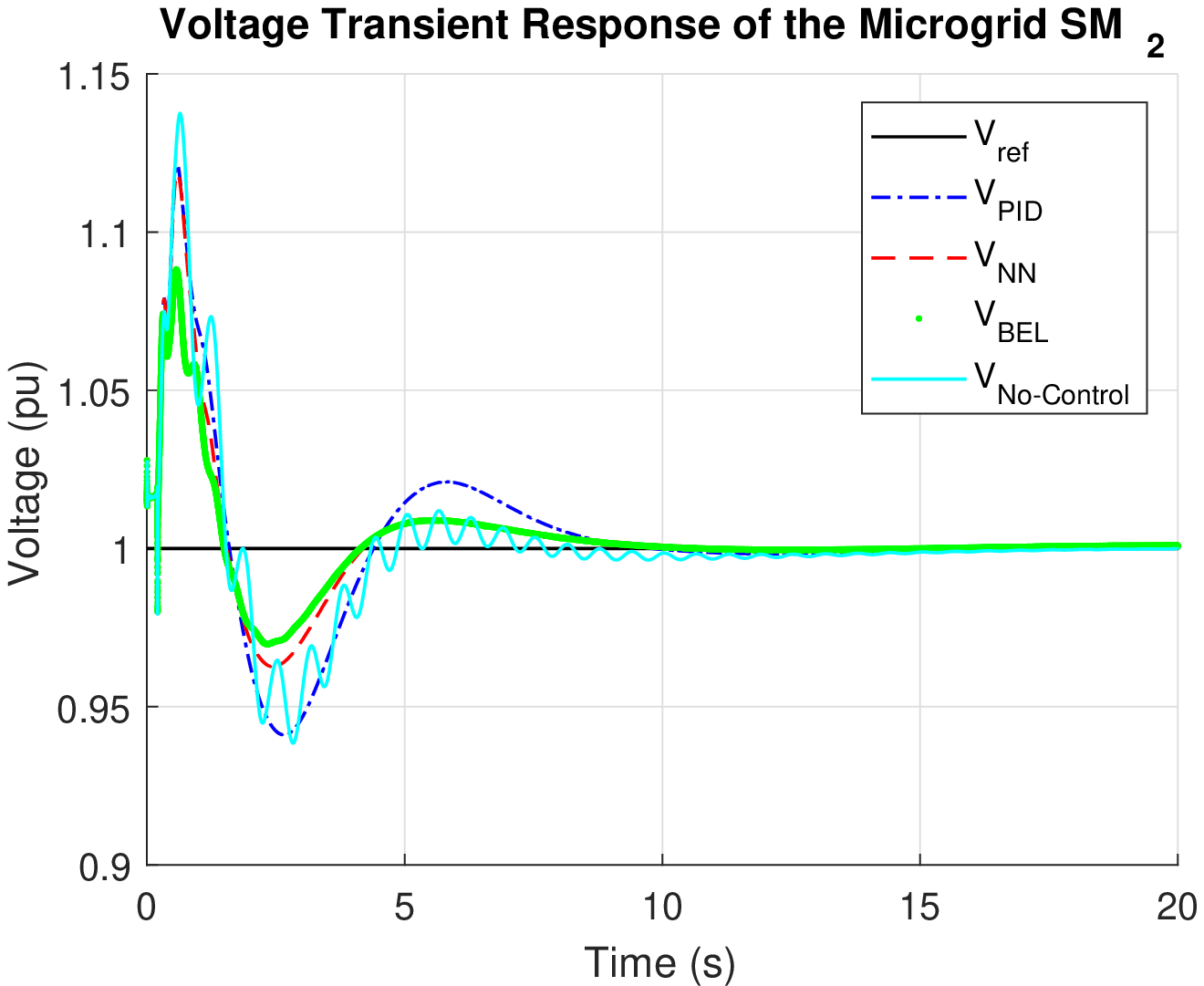}
\includegraphics[width=0.5\columnwidth]{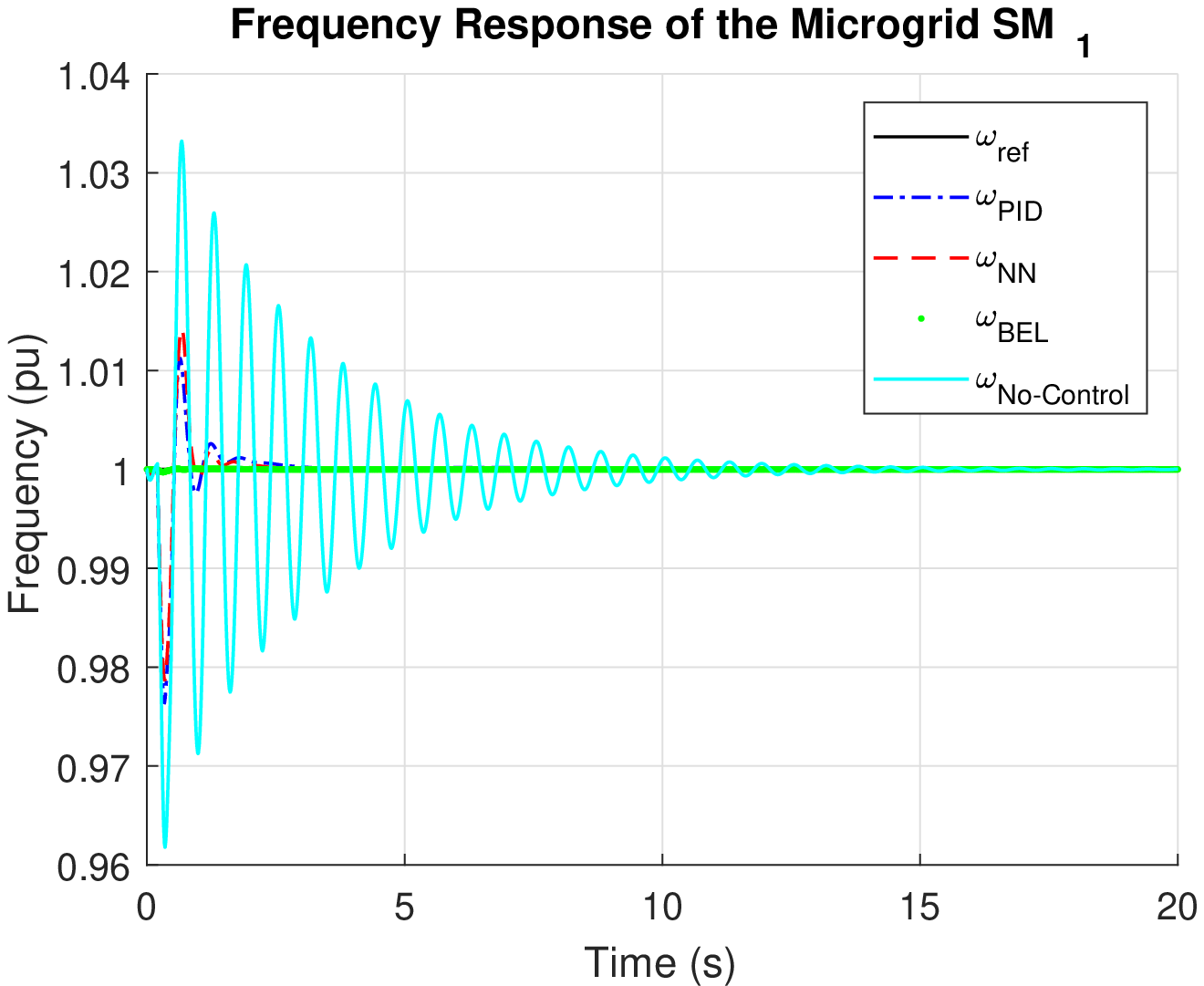} 
\includegraphics[width=0.5\columnwidth]{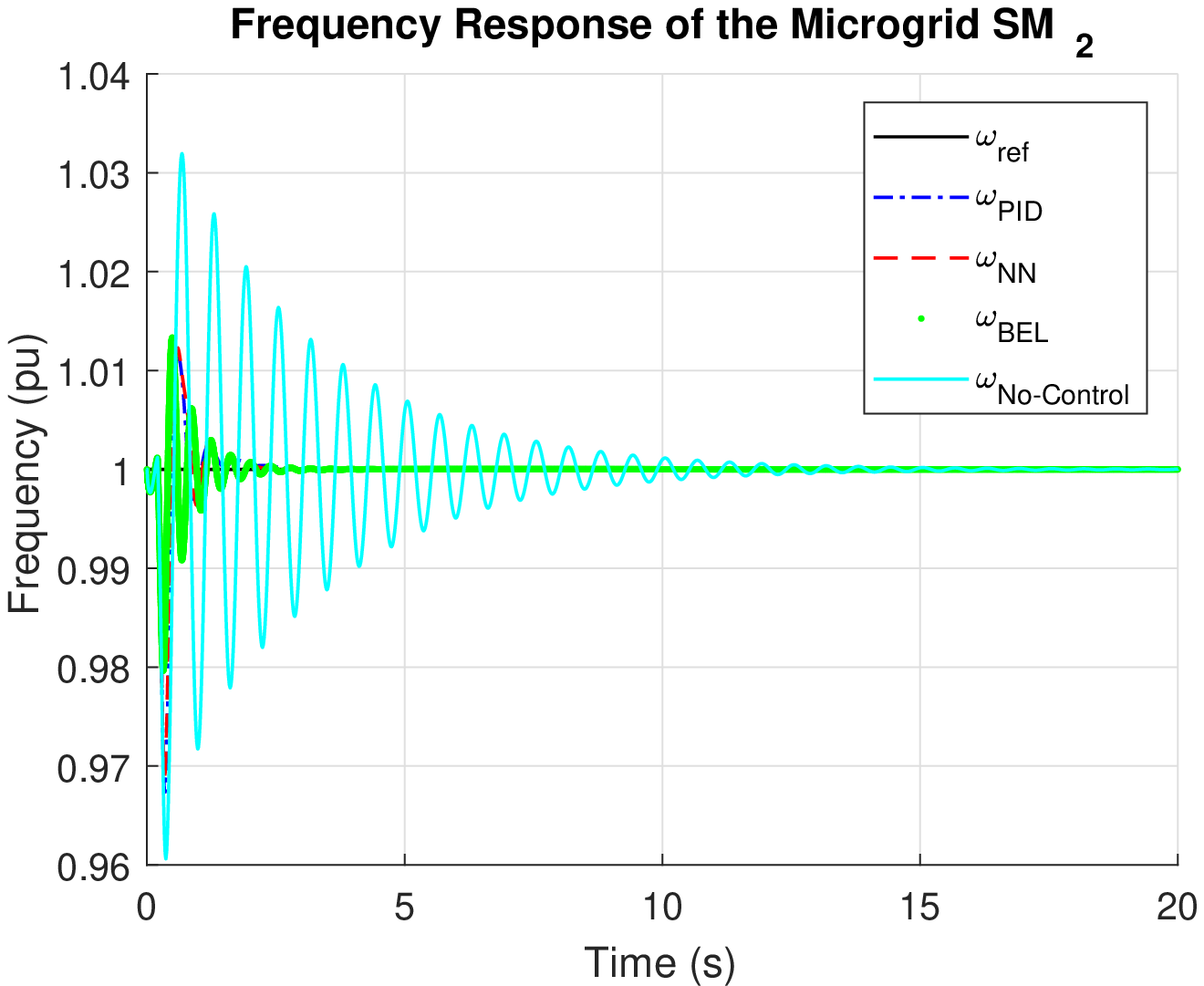}
\end{array}$
\end{center}
\vspace{-15pt}
    \caption{The frequency and voltage outputs of the system ($SM_1$ and $SM_2$) for the proposed intelligent secondary control. In all cases, BELBIC-based intelligent controller is in Green color, NN-based intelligent controller is in Red color, the PID controller is in Blue color, the system with no secondary controller is in Cyan color, and the reference signal is shown in Black. MG will be disconnected from the main grid at $t=0.2$ second.}
    \label{fig:results1}
\vspace{-10pt}
\end{figure*}

\begin{figure*}[htb!]
\begin{center}$
\begin{array}{cccc}
\includegraphics[width=0.5\columnwidth]{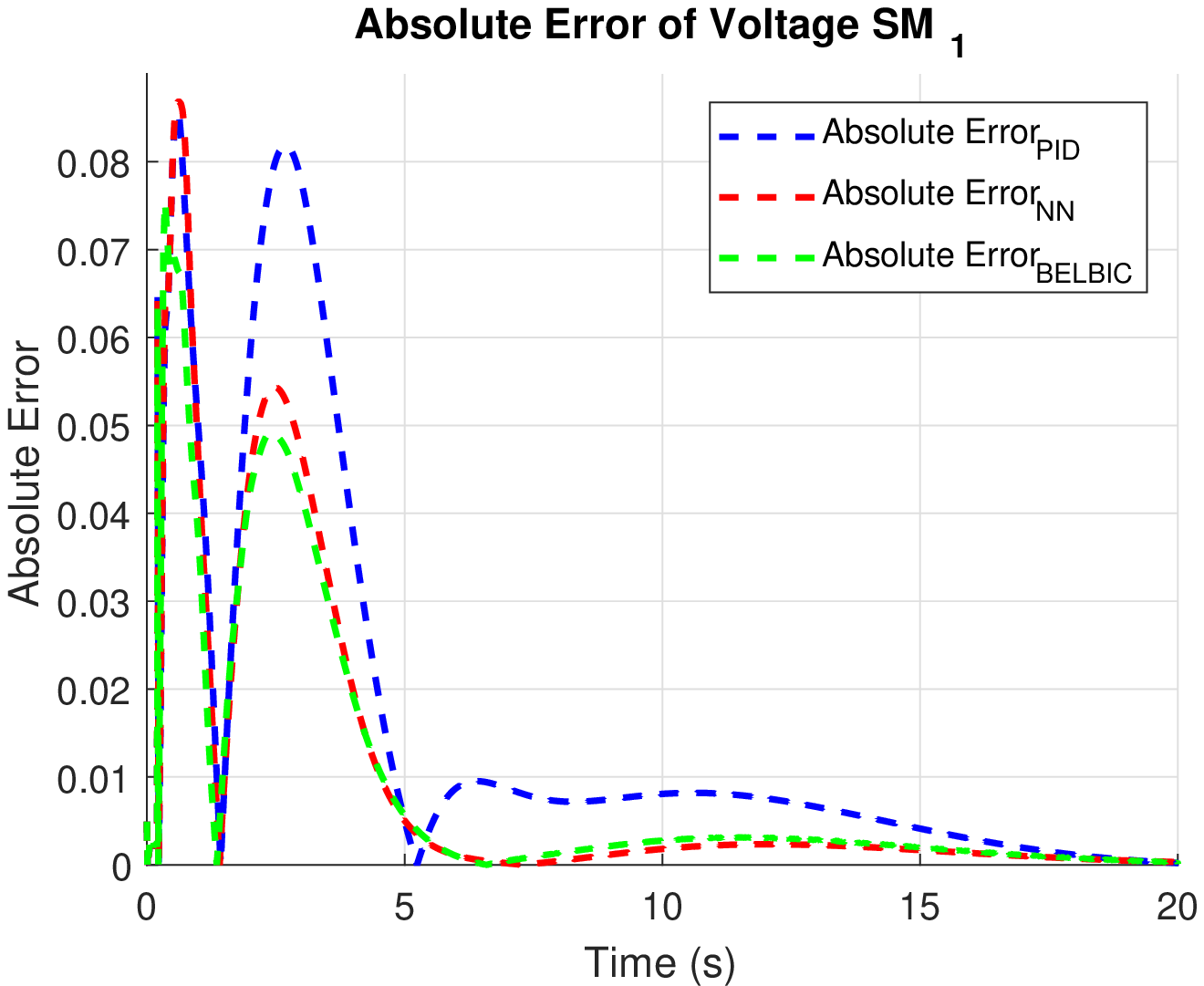} 
\includegraphics[width=0.5\columnwidth]{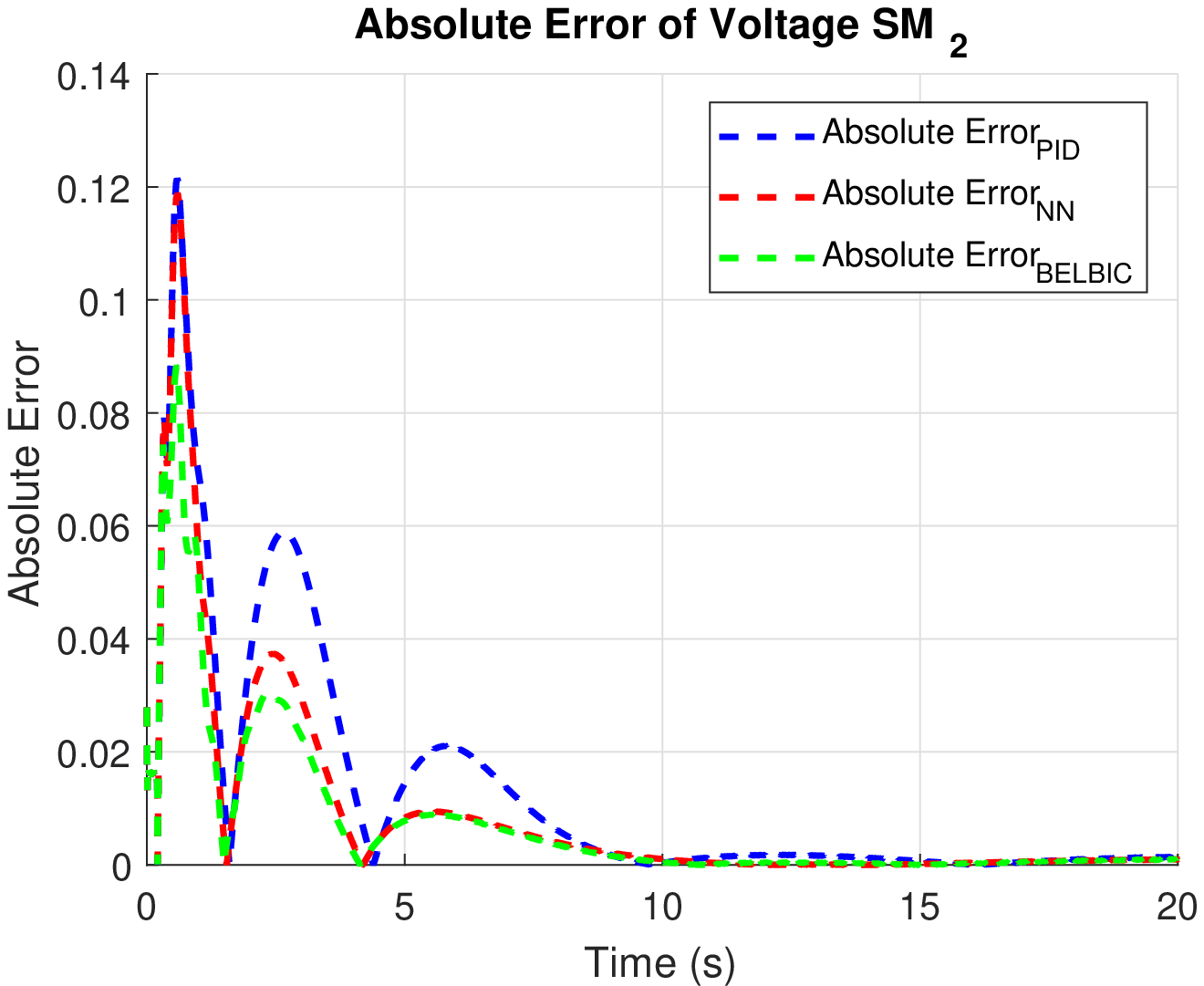}
\includegraphics[width=0.5\columnwidth]{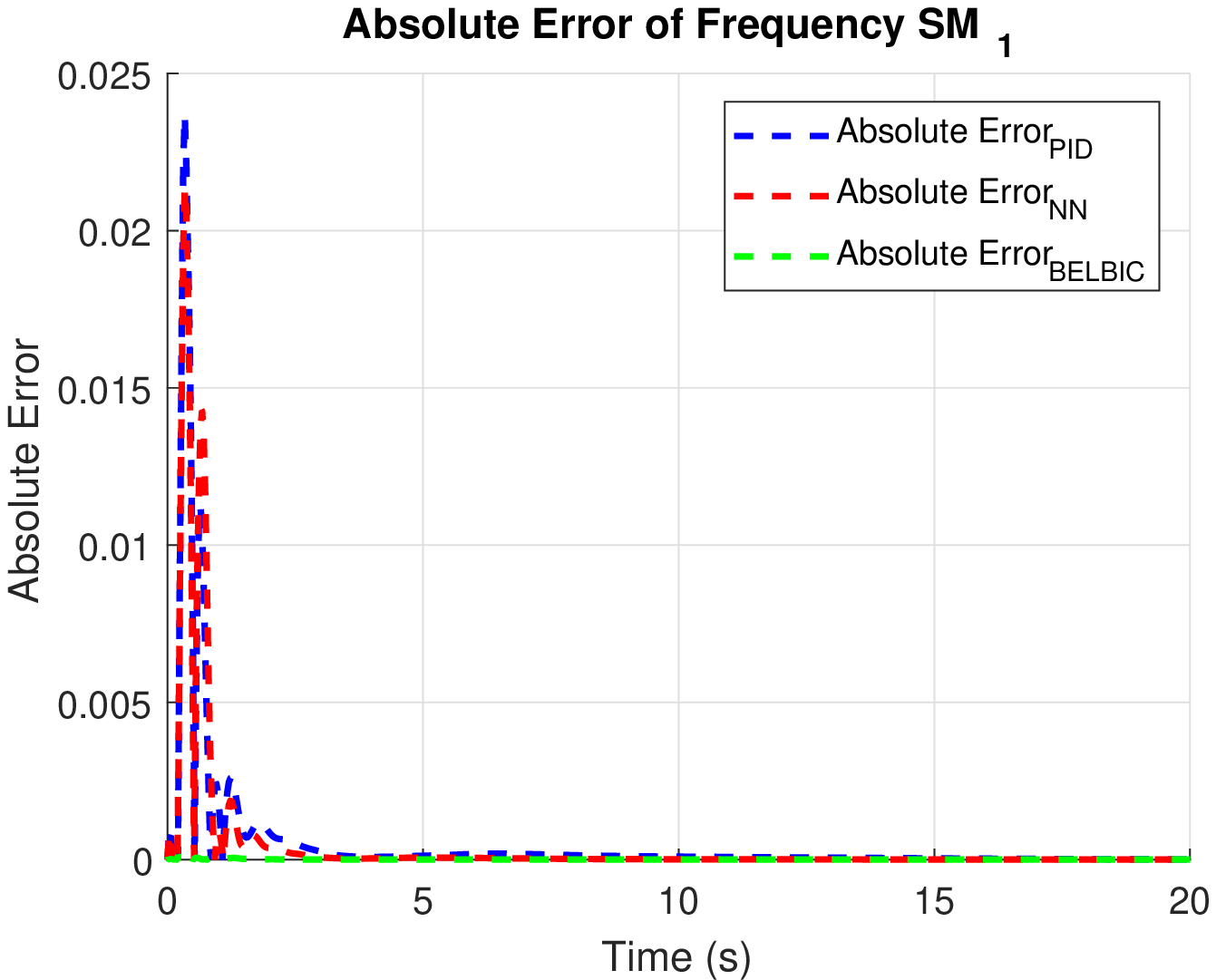} 
\includegraphics[width=0.5\columnwidth]{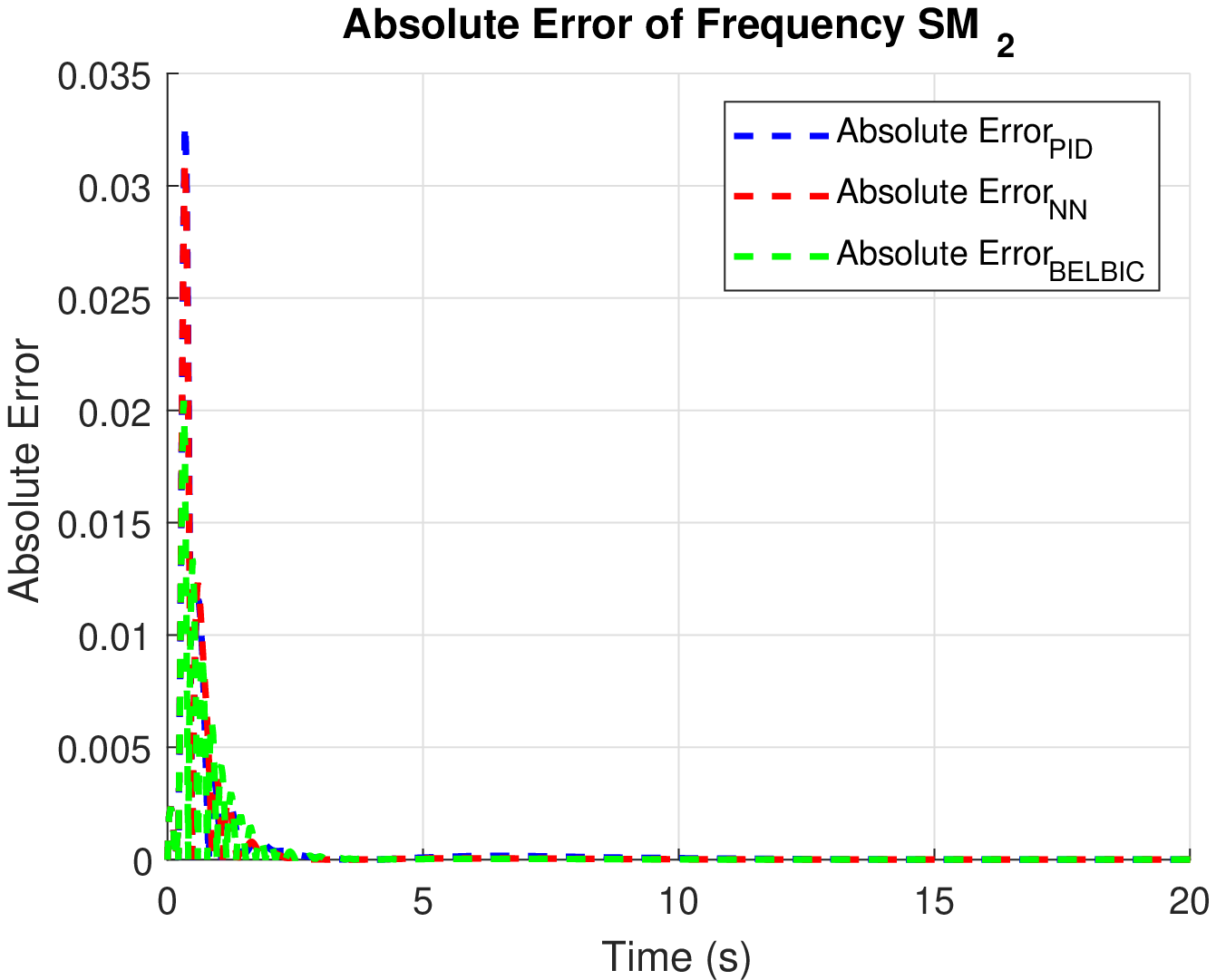}
\end{array}$
\end{center}
\vspace{-15pt}
    \caption{The Frequency and Voltage Absolute Error of the system ($SM_1$ and $SM_2$) for all secondary controllers. BELBIC-based intelligent controller is in Green color, NN-based intelligent controller is in Red color, and the conventional PID controller is in Blue color.}
    \label{fig:mse_all}
\vspace{-10pt}
\end{figure*}
\subsection{Case Study}
This section presents the simulation results of the MG consisting of two synchronous generators, two asynchronous motors, loads and lines under the transition from the grid-connected mode to islanded mode. In this MG the $SM_1$ is operating as a slack generator and the secondary controller is only providing the control inputs of $SM_1$. To get a better insight regarding the system model, a differential and algebraic equations (DAEs) scheme is developed to design the PID controller which is then used as the benchmark controller. The single line diagram of the sample MG model and detailed explanation of the system model is provided in \cite{JafariBELBIC2018}. 

In this paper, simulation results are provided under four different scenarios: no secondary controller, simple PID secondary controller, NN-based secondary controller~\cite{jafari2018power}, and BELBIC-based secondary controller. In the first scenario, there is no secondary control and values of $U_{sec}$ and $T_{sec}$ are equal to zero. In the second scenario, two separate PID controller are responsible for generating $U_{sec}$ and $T_{sec}$ in each time step. In the third scenario, NN-based secondary controller is employed to generate $U_{sec}$ and $T_{sec}$ for the SM. Finally, in the last scenario BELBIC-based secondary controller is utilized to generate $U_{sec}$ and $T_{sec}$ for the SM. In all scenarios, the total simulation time is 20 seconds and MG is disconnected from the main grid at $t=0.2$ second. Also, the sampling time for the all simulations is $0.001$s. 
All simulations are carried out on a Win 7 platform with CPU of E5420 2*2.5 GHz and 16 GB of RAM.
\subsection{Simulation Results}
Fig.~\ref{fig:results1} shows the frequency response and the voltage transient of the MG for $SM_1$ and $SM_2$ during the transition period. In all cases, BELBIC-based intelligent controller is shown in Green color, NN-based intelligent controller is shown in Red color, the conventional PID controller is plotted in Blue color, the system with no secondary controller is shown in Cyan color, and the reference signal is in Black color. From these plots, it is observed that all the controllers, i.e., PID, NN-based and the proposed BELBIC-based methods are capable of stabilizing both the frequency and the voltage. However, the  proposed BELBIC-based controller have faster response, lower overshoot, smaller settling time in comparison with both PID and NN-based controllers. Furthermore, when the MG has been disconnected from the main grid at 0.2 second, the proposed method responds faster than the other methods to successfully handle the sudden changes in the system. 

Fig.~\ref{fig:mse_all} shows the absolute error of the frequency and voltage output of $SM_1$ and $SM_2$ for all secondary controllers. By comparing the voltage and frequency absolute errors of the proposed BELBIC-based controller with both the PID and NN-based controllers, it is observed that the proposed method can compensate both voltage and frequency with less error. Therefore, it is more appropriate for controlling the MG system. 

\subsection{Sensitivity Analysis}
\begin{figure*}[htb!]
\begin{center}$
\begin{array}{cccc}
\includegraphics[width=0.5\columnwidth]{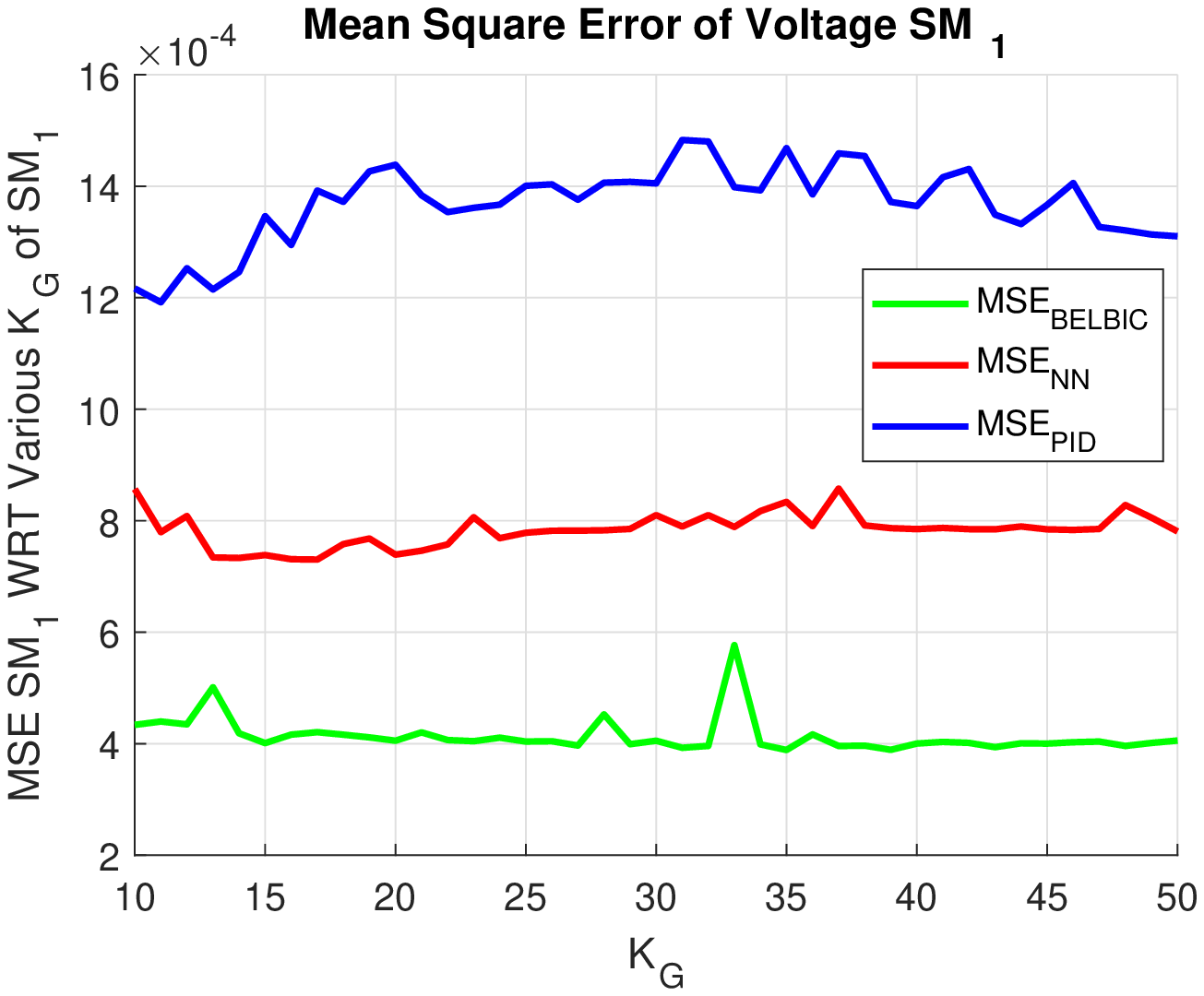} 
\includegraphics[width=0.5\columnwidth]{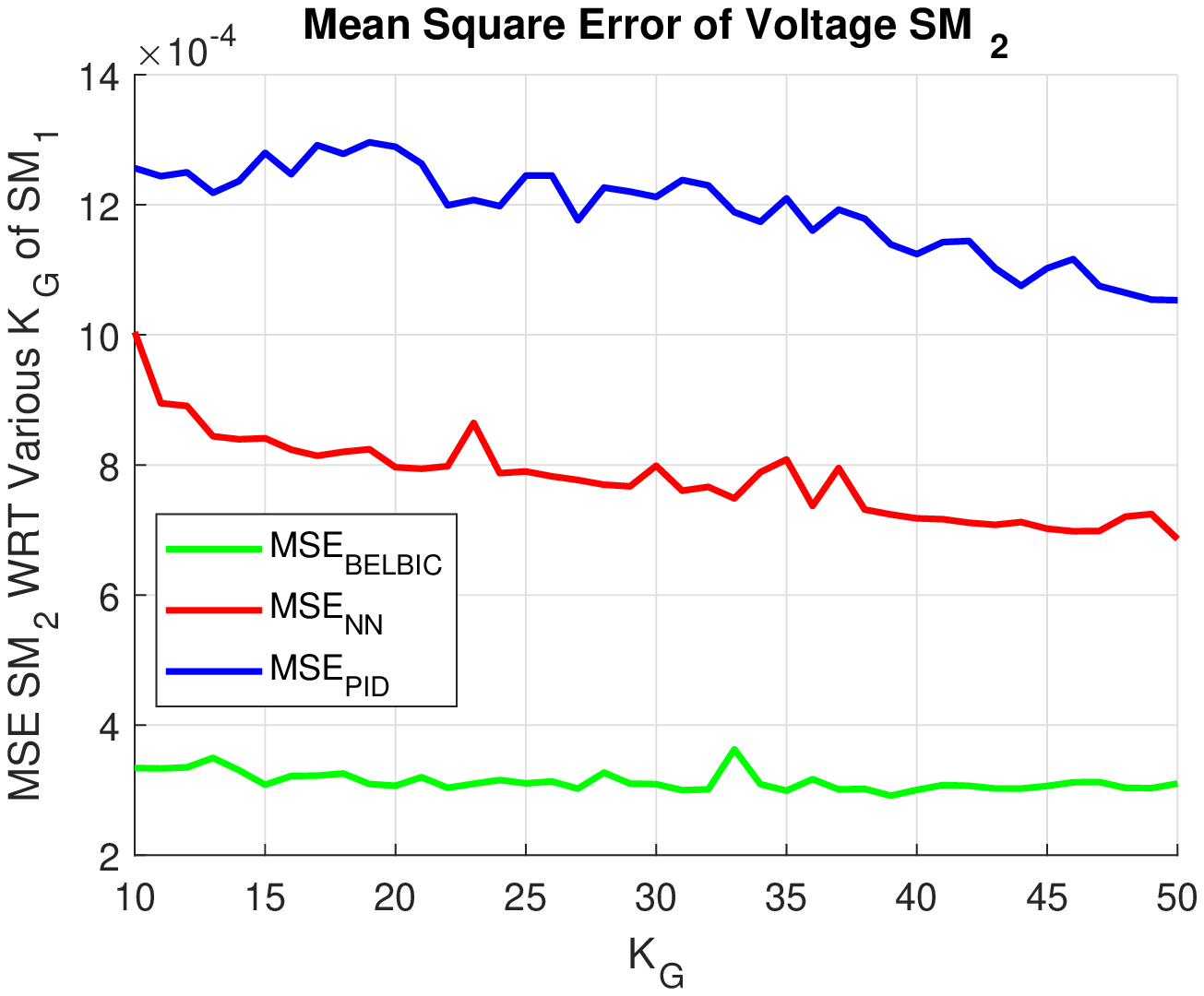}
\includegraphics[width=0.5\columnwidth]{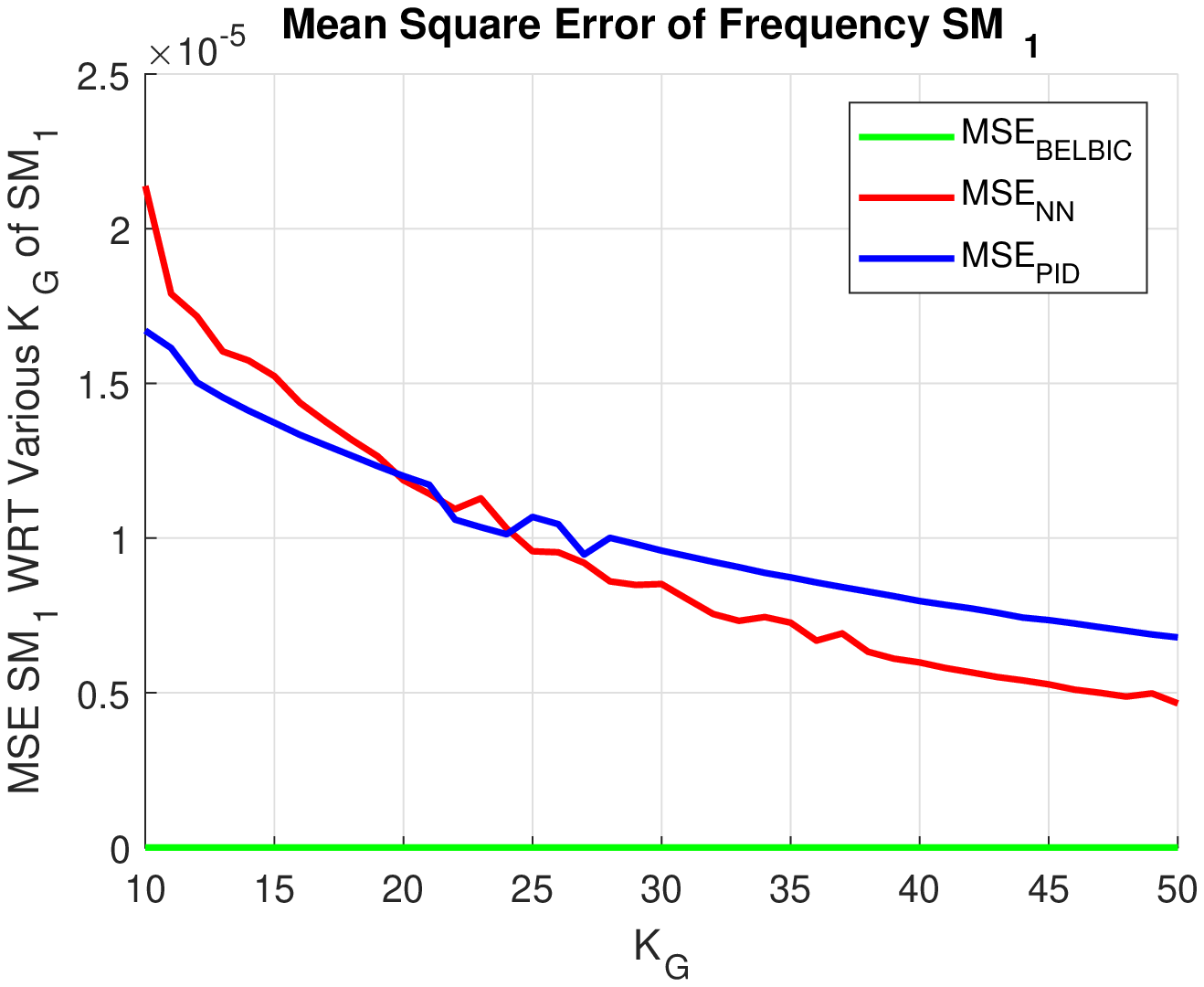} 
\includegraphics[width=0.5\columnwidth]{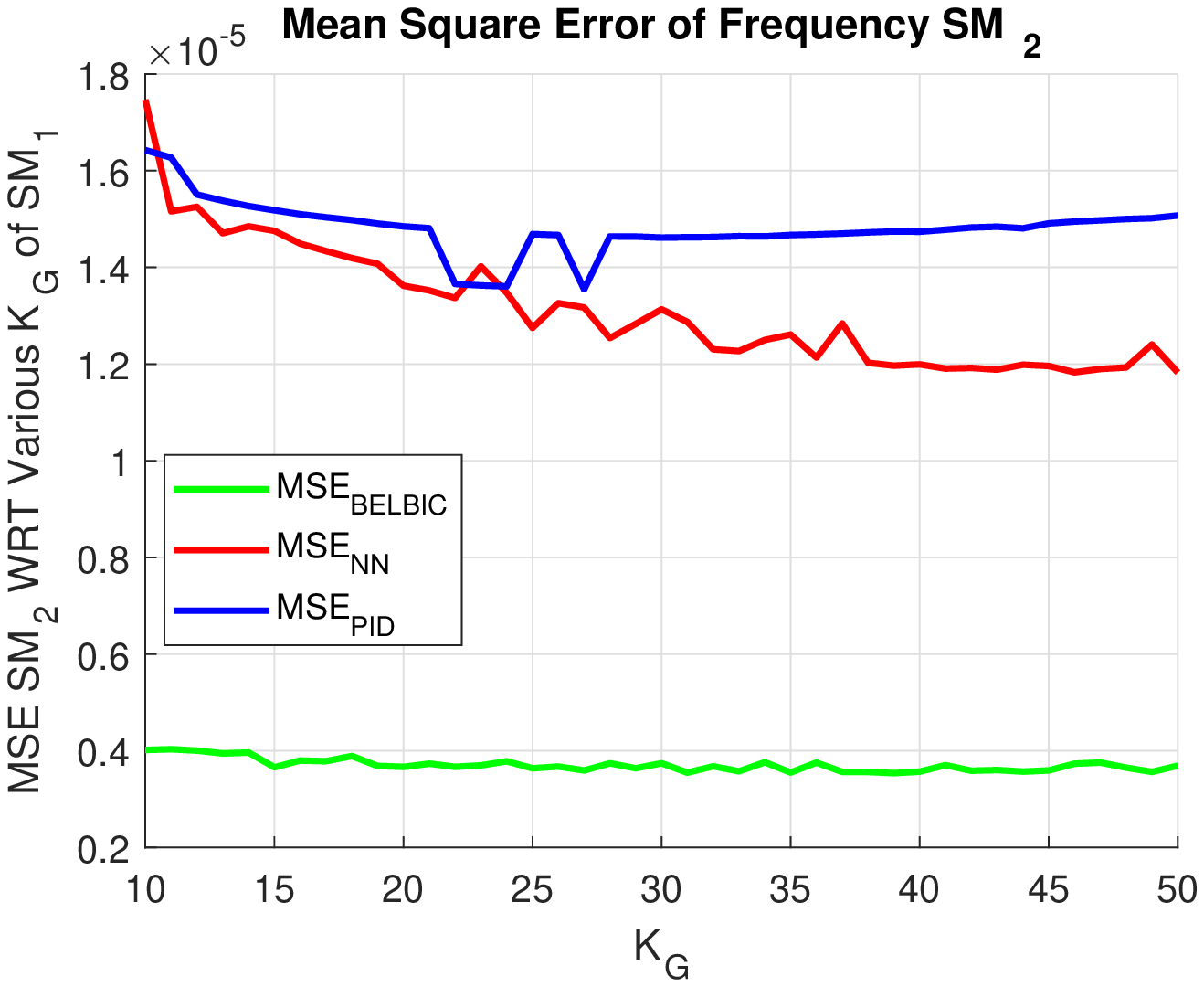}
\end{array}$
\end{center}
\vspace{-15pt}
    \caption{The Frequency and Voltage Mean Square Error of $SM_1$ and $SM_2$ for all secondary controllers considering the different values for Turbine gain ($K_G$) in $SM_1$. BELBIC-based intelligent controller is in Green color, NN-based intelligent controller is in Red color, the PID controller is in Blue color.}
    \label{fig:Mmse_all}
\vspace{-10pt}
\end{figure*}
\begin{figure*}[htb!]
\begin{center}$
\begin{array}{cccc}
\includegraphics[width=0.5\columnwidth]{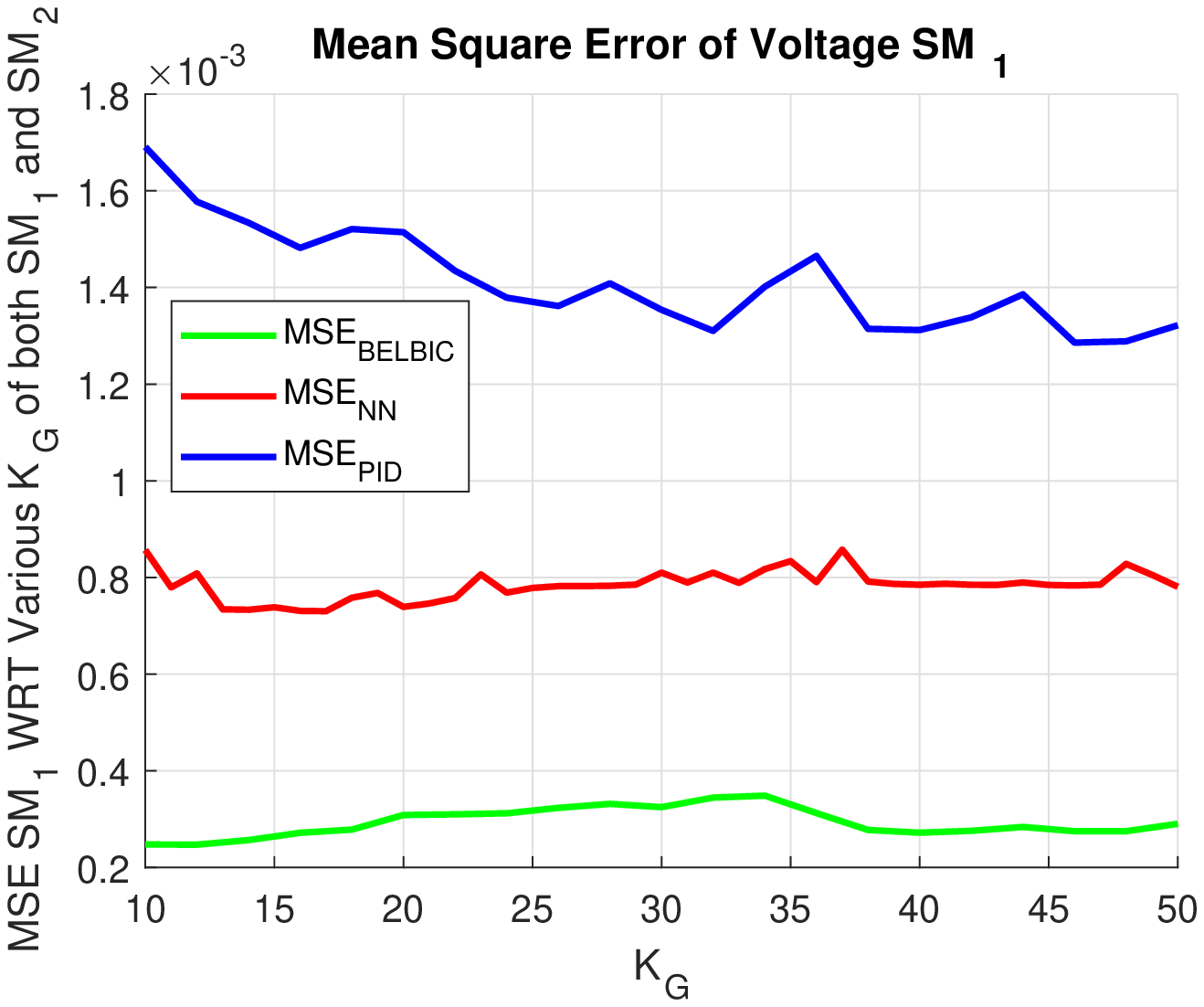} 
\includegraphics[width=0.5\columnwidth]{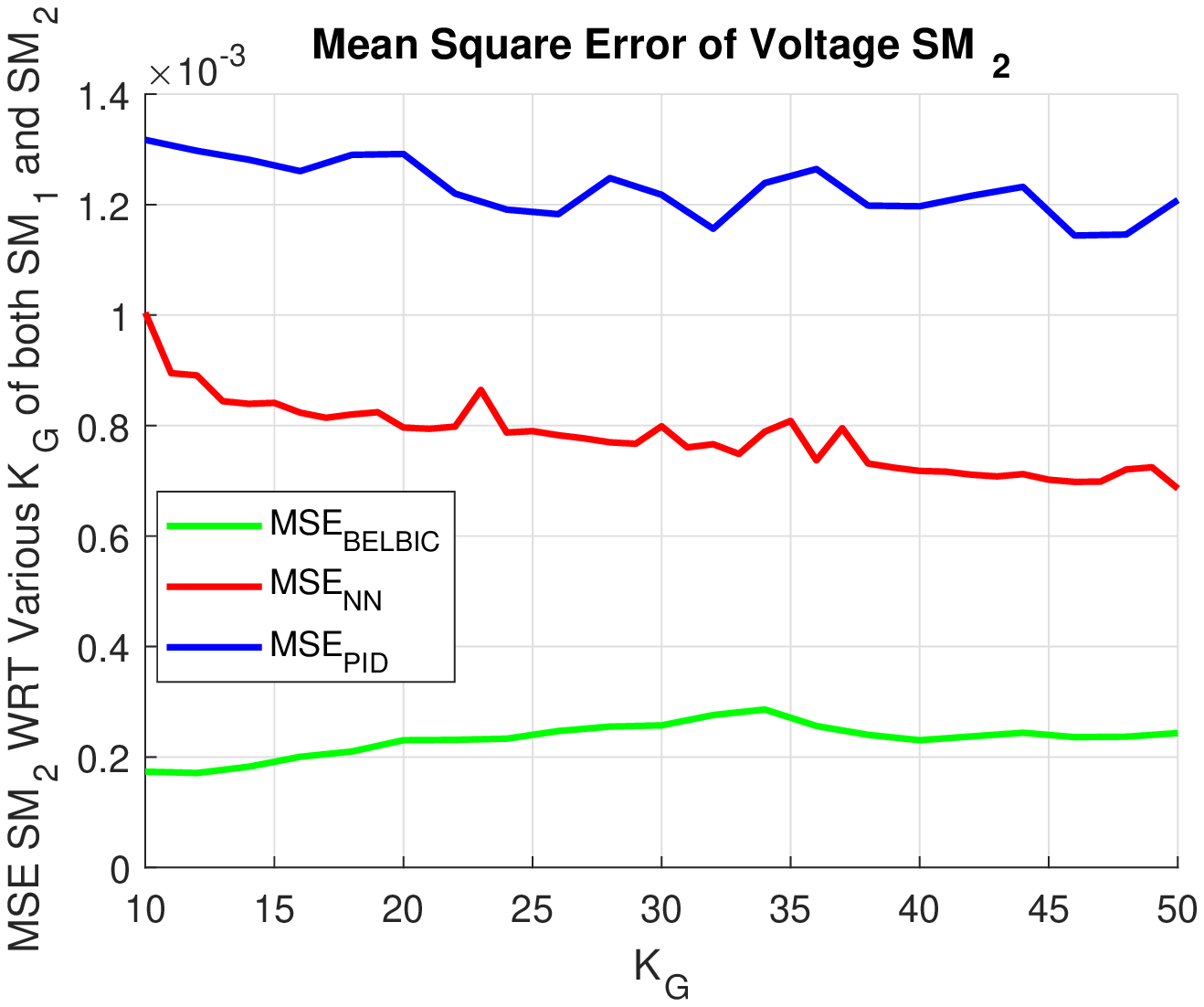}
\includegraphics[width=0.5\columnwidth]{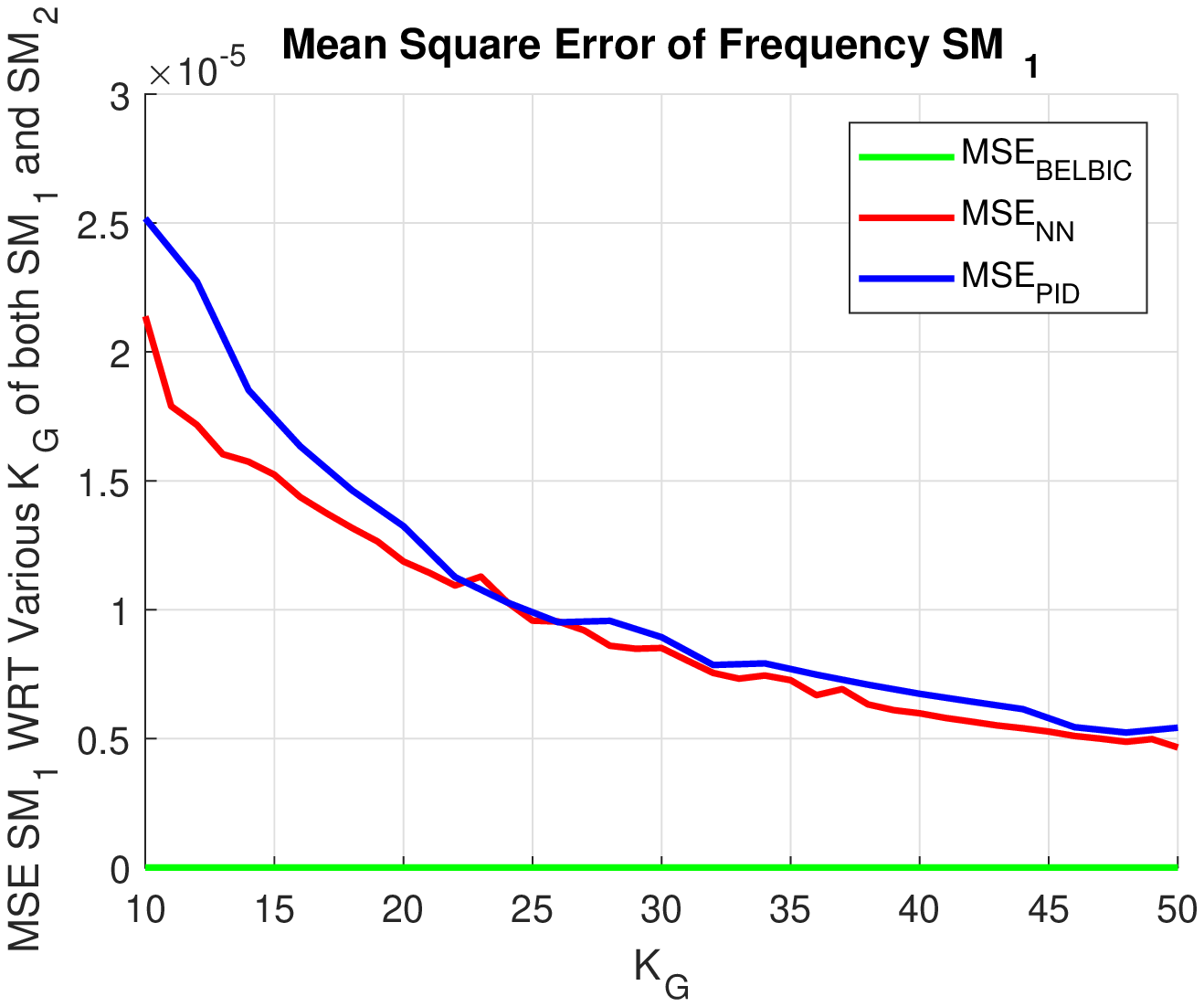} 
\includegraphics[width=0.5\columnwidth]{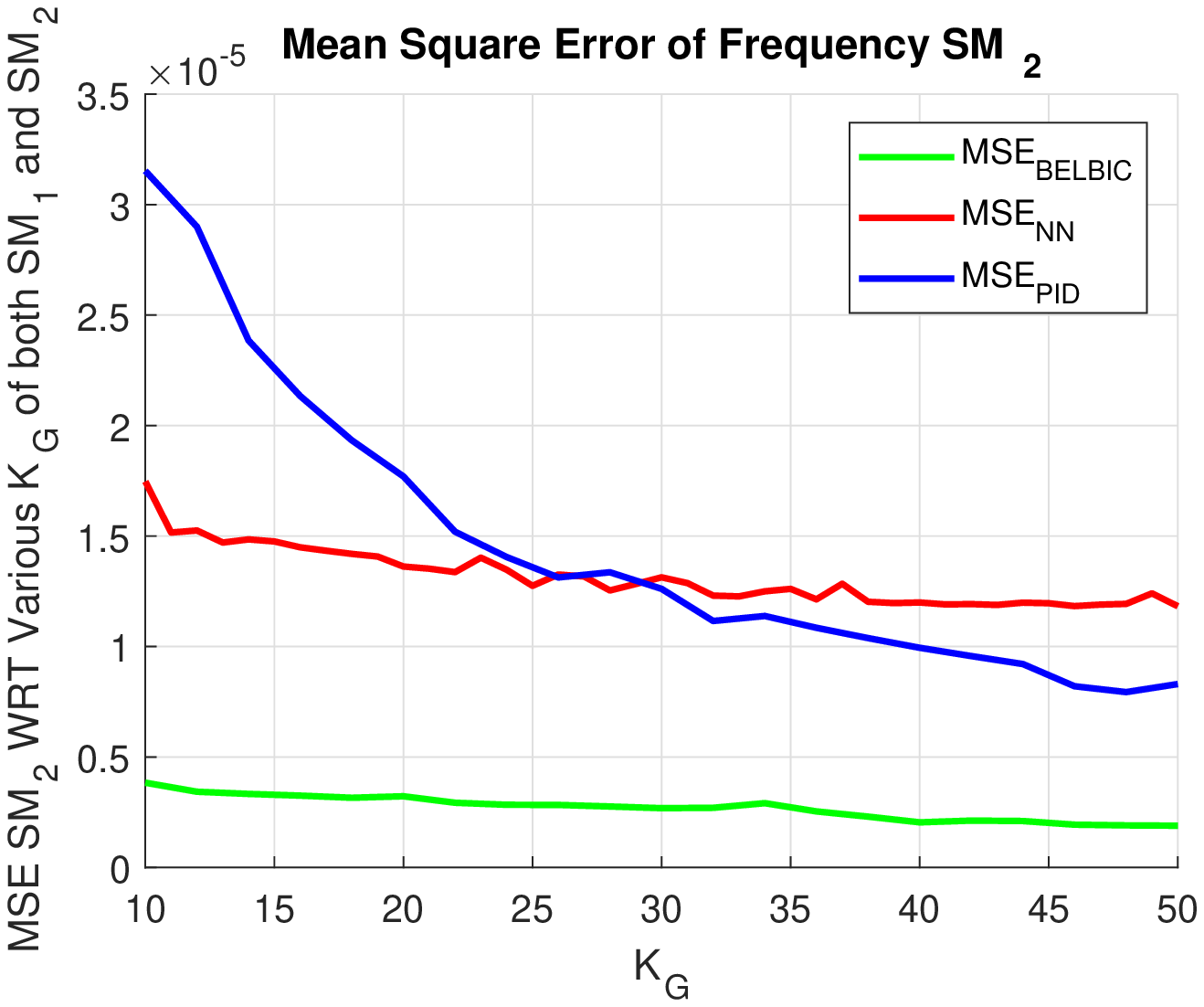}
\end{array}$
\end{center}
\vspace{-15pt}
    \caption{The Frequency and Voltage Mean Square Error of $SM_1$ and $SM_2$ for all secondary controllers considering the different values for Turbine gain ($K_G$) in both $SM_1$ and $SM_2$. BELBIC-based intelligent controller is in Green color, NN-based intelligent controller is in Red color, the PID controller is in Blue color.}
    \label{fig:Mmse_all2}
\vspace{-20pt}
\end{figure*}

Aside from faults and/or disturbances, the dynamic of the system might be fully or partially unknown and even after employing identification methods, the characteristics of the system might change due to wear and tear, environmental changes, etc. Therefore, different simulations have been done to study the sensitivity of all controllers with respect to variations in system parameters such as: AVR time constant of the regulator ($T_{a}$), AVR regulator gain ($K_{a}$), Turbine gain ($K_{G}$), Turbine time constant ($T_{G}$). The objective is to evaluate the performance of all secondary controllers, considering that their settings remain the same as before. In other words, there was no additional tuning of the controllers for adapting to the new system parameters. For the sake of brevity, only the results for the variations in the turbine gain is discussed.

Fig.~\ref{fig:Mmse_all}  shows the frequency and voltage Mean Square Error (MSE)  of $SM_1$ and $SM_2$ for all secondary controllers considering the different values for turbine gain ($K_G$) in $SM_1$. In addition, Fig.~\ref{fig:Mmse_all2} shows the frequency and voltage Mean Square Error (MSE)  of $SM_1$ and $SM_2$ for all secondary controllers considering the different values for turbine gain ($K_G$) in both $SM_1$ and $SM_2$. These figures demonstrate that the proposed method has both less MSE and small variation due to the changing of system parameters. Therefore, it is more robust and appropriate for controlling the MG system.

\section{Conclusion and Future Work}
\label{section:conclusions}
A biologically-inspired adaptive intelligent secondary controller for microgrids is developed in this paper. The proposed methodology is designed and implemented for stabilizing the frequency and the voltage of a microgrid in presence of system dynamics uncertainties and disturbances. The simulation results demonstrate the effectiveness of the proposed method.

In the future work, the proposed secondary controller will be extended to multiple community-based microgrids through an independent operator, called community-based microgrid operator (C-MGO). In addition, multiple community-based microgrids will be implemented in RTDS platform to simulate different disturbances and uncertainties in a real-time system.



\begin{thebibliography}{10}
\providecommand{\url}[1]{#1}
\csname url@samestyle\endcsname
\providecommand{\newblock}{\relax}
\providecommand{\bibinfo}[2]{#2}
\providecommand{\BIBentrySTDinterwordspacing}{\spaceskip=0pt\relax}
\providecommand{\BIBentryALTinterwordstretchfactor}{4}
\providecommand{\BIBentryALTinterwordspacing}{\spaceskip=\fontdimen2\font plus
\BIBentryALTinterwordstretchfactor\fontdimen3\font minus
  \fontdimen4\font\relax}
\providecommand{\BIBforeignlanguage}[2]{{%
\expandafter\ifx\csname l@#1\endcsname\relax
\typeout{** WARNING: IEEEtran.bst: No hyphenation pattern has been}%
\typeout{** loaded for the language `#1'. Using the pattern for}%
\typeout{** the default language instead.}%
\else
\language=\csname l@#1\endcsname
\fi
#2}}
\providecommand{\BIBdecl}{\relax}
\BIBdecl

\bibitem{bidram2014distributed}
A.~Bidram, F.~L. Lewis, and A.~Davoudi, ``Distributed control systems for
  small-scale power networks: Using multiagent cooperative control theory,''
  \emph{IEEE Control Systems}, vol.~34, no.~6, pp. 56--77, 2014.

\bibitem{bevrani2012intelligent}
H.~Bevrani, F.~Habibi, P.~Babahajyani, M.~Watanabe, and Y.~Mitani,
  ``Intelligent frequency control in an ac microgrid: Online pso-based fuzzy
  tuning approach,'' \emph{IEEE Transactions on Smart Grid}, vol.~3, no.~4, pp.
  1935--1944, 2012.

\bibitem{lasseter2011certs}
R.~H. Lasseter, J.~H. Eto, B.~Schenkman, J.~Stevens, H.~Vollkommer, D.~Klapp,
  E.~Linton, H.~Hurtado, and J.~Roy, ``Certs microgrid laboratory test bed,''
  \emph{IEEE Transactions on Power Delivery}, vol.~26, no.~1, pp. 325--332,
  2011.

\bibitem{mohamed2017hierarchical}
A.~A. Mohamed, A.~T. Elsayed, T.~A. Youssef, and O.~A. Mohammed, ``Hierarchical
  control for dc microgrid clusters with high penetration of distributed energy
  resources,'' \emph{Electric Power Systems Research}, vol. 148, pp. 210--219,
  2017.

\bibitem{katiraei2005micro}
F.~Katiraei, M.~R. Iravani, and P.~W. Lehn, ``Micro-grid autonomous operation
  during and subsequent to islanding process,'' \emph{IEEE Transactions on
  power delivery}, vol.~20, no.~1, pp. 248--257, 2005.

\bibitem{manaffam2017intelligent}
S.~Manaffam, M.~Talebi, A.~Jain, and A.~Behal, ``Intelligent pinning based
  cooperative secondary control of distributed generators for microgrid in
  islanding operation mode,'' \emph{IEEE Transactions on Power Systems}, 2017.

\bibitem{lou2017decentralised}
G.~Lou, W.~Gu, L.~Wang, B.~Xu, M.~Wu, and W.~Sheng, ``Decentralised secondary
  voltage and frequency control scheme for islanded microgrid based on adaptive
  state estimator,'' \emph{IET Generation, Transmission \& Distribution},
  vol.~11, no.~15, pp. 3683--3693, 2017.

\bibitem{li2018distributed}
Q.~Li, C.~Peng, M.~Wang, M.~Chen, J.~M. Guerrero, and D.~Abbott, ``Distributed
  secondary control and management of islanded microgrids via dynamic
  weights,'' \emph{IEEE Transactions on Smart Grid}, 2018.

\bibitem{bevrani2013intelligent}
H.~Bevrani and S.~Shokoohi, ``An intelligent droop control for simultaneous
  voltage and frequency regulation in islanded microgrids,'' \emph{IEEE
  transactions on smart grid}, vol.~4, no.~3, pp. 1505--1513, 2013.

\bibitem{jafari2013attitude}
M.~Jafari, A.~M. Shahri, and S.~B. Shouraki, ``Attitude control of a quadrotor
  using brain emotional learning based intelligent controller,'' in \emph{Fuzzy
  Systems (IFSC), 2013 13th Iranian Conference on}.\hskip 1em plus 0.5em minus
  0.4em\relax IEEE, 2013, pp. 1--5.

\bibitem{klecker2017robust}
S.~Klecker, B.~Hichri, and P.~Plapper, ``Robust belbic-extension for trajectory
  tracking control,'' \emph{Journal of Mechanics Engineering and Automation},
  vol.~7, no.~2, 2017.

\bibitem{yin2017artificial}
L.~Yin, T.~Yu, L.~Zhou, L.~Huang, X.~Zhang, and B.~Zheng, ``Artificial
  emotional reinforcement learning for automatic generation control of
  large-scale interconnected power grids,'' \emph{IET Generation, Transmission
  \& Distribution}, vol.~11, no.~9, pp. 2305--2313, 2017.

\bibitem{jafari2018power}
M.~Jafari, V.~Sarfi, A.~Ghasemkhani, H.~Livani, L.~Yang, H.~Xu, and R.~Koosha,
  ``Adaptive neural network based intelligent secondary control for
  microgrids,'' in \emph{Texas Power and Energy Conference (TPEC), 2018
  IEEE}.\hskip 1em plus 0.5em minus 0.4em\relax IEEE, 2018, pp. 1--6.

\bibitem{lucas2004introducing}
C.~Lucas, D.~Shahmirzadi, and N.~Sheikholeslami, ``Introducing belbic: brain
  emotional learning based intelligent controller,'' \emph{Intelligent
  Automation \& Soft Computing}, vol.~10, no.~1, pp. 11--21, 2004.

\bibitem{jafari2017mas}
M.~Jafari, H.~Xu, and L.~R.~G. Carrillo, ``Brain emotional learning-based
  intelligent controller for flocking of multi-agent systems,'' in
  \emph{American Control Conference (ACC), 2017}.\hskip 1em plus 0.5em minus
  0.4em\relax IEEE, 2017, pp. 1996--2001.

\bibitem{kim2017brain}
J.-W. Kim, C.-Y. Oh, J.-W. Chung, and K.-H. Kim, ``Brain emotional limbic-based
  intelligent controller design for control of a haptic device,''
  \emph{International Journal of Automation and Control}, vol.~11, no.~4, pp.
  358--371, 2017.

\bibitem{moren2000computational}
J.~Moren and C.~Balkenius, ``A computational model of emotional learning in the
  amygdala,'' \emph{From animals to animats}, vol.~6, pp. 115--124, 2000.

\bibitem{JafariBELBIC2018}
\BIBentryALTinterwordspacing
``Technical report: Adaptive intelligent secondary control of microgrids using
  a biologically-inspired reinforcementlearning.'' [Online]. Available:
  \url{TBD}
\BIBentrySTDinterwordspacing

\end{thebibliography}
\end{document}